\documentclass[aps,prc,onecolumn,superscriptaddress,showpacs,showkeys,nofootinbib]{revtex4-2}
\usepackage{graphicx}
\usepackage{bm}
\usepackage{epsfig}
\usepackage{amsfonts}
\usepackage{amssymb,amscd}
\usepackage{subfigure}
\usepackage{xcolor}
\usepackage{comment}
\usepackage{amsmath}
\usepackage{graphicx}
\usepackage{dcolumn}
\usepackage{bm}
\usepackage{hyperref}

\newcommand{\rb}{\mbox{\boldmath $b$}}

\begin{document}


\title{Inclusive heavy meson photoproduction in $pPb$ and $PbPb$ collisions}

\author{Victor P. {\sc Gon\c{c}alves}}
\email{barros@ufpel.edu.br}
\affiliation{Institute of Physics and Mathematics, Federal University of Pelotas, Postal Code 354,  96010-900, Pelotas, RS, Brazil}

\author{Luana {\sc Santana}}
\email{luanas1899@gmail.com}
\affiliation{Institute of Physics and Mathematics, Federal University of Pelotas, Postal Code 354,  96010-900, Pelotas, RS, Brazil}

\author{Wolfgang {\sc Schäfer}}
\email{Wolfgang.Schafer@ifj.edu.pl}
\affiliation{The Henryk Niewodniczański Institute of Nuclear Physics
Polish Academy of Sciences (PAN), 31-342 Kraków, Poland.}

\date{\today}

\begin{abstract}
The inclusive photoproduction of heavy mesons in ultraperipheral $pPb$ and $PbPb$ collisions at the LHC energies is investigated considering the color dipole $S$-matrix formalism and assuming distinct models for the unintegrated gluon distribution, based on different assumptions for the description of the QCD dynamics. In particular, predictions for the $B^0$-meson photoproduction are presented here for the  first time. The study of the $D^0$-meson photoproduction is revisited by estimating the impact of the treatment of the heavy charm fragmentation on the predictions and extended for $pPb$ collisions. Moreover, the contribution associated with the  $b \rightarrow D^0$ transition is estimated. Our results indicate that a future experimental analysis of the heavy meson photoproduction will provide important constrains on the description of the hadronic structure at high energies.   
\end{abstract}

\maketitle


\section{Introduction}
Heavy meson production in high energy photon - induced interactions is a prominent testing ground for various perturbative QCD approaches. In particular, such a process is expected to provide important constrains on the target gluon distribution as well on the description of the QCD dynamics at high energies (small values of the Bjorken - $x$ variable). During the last decades, the photoproduction of heavy vector mesons at the LHC in exclusive processes, where both incident particles remain intact,  was largely investigated (for reviews see Ref.~\cite{upc}) and more precise data are expected  in the forthcoming years. In contrast, experimental results for the  heavy meson photoproduction in inclusive processes, where one of the incident hadrons breakup, were only recently released 
\cite{CMS:2025jjx,Nese:2025ohz}, which  have motivated the improvement of the theoretical description of this process performed in Refs. \cite{Gimeno-Estivill:2025rbw,Goncalves:2025wwt,Cacciari:2025tgr}    \footnote{For related studies about inclusive photon - induced processes in hadronic collisions, see  e.g. Refs. \cite{Klein:2000dk,Klein:2002wm,Goncalves:2003is,Goncalves:2004dn,Goncalves:2009ey,Goncalves:2013oga,Goncalves:2015cik,Kotko:2017oxg,Goncalves:2017zdx,Guzey:2019kik,Guzey:2018dlm,Eskola:2024fhf,Goncalves:2019owz}).}.

Motivated by the perspective of the releasing of more precise data in the coming years, our goal in this paper is twofold: (a) update the predictions for the inclusive $D^0$ photoproduction in $PbPb$ performed in Ref.~\cite{Goncalves:2025wwt}, by considering a fragmentation function that evolves with the hard scale;   and (b) present, for the first time, predictions for $pPb$ collisions and for the inclusive photoproduction of $B^0$ mesons. Our calculations will be performed using the  color dipole $S$-matrix formalism \cite{Nikolaev:2003zf,Nikolaev:2004cu,Nikolaev:2005dd,Nikolaev:2005zj,Nikolaev:2005ay,Nikolaev:2005qs}, considering different models for the target unintegrated gluon distribution, based on distinct assumptions for the description of the QCD dynamics, and different magnitudes for the electromagnetic dissociation. As we will demonstrate below, our results indicate that a future experimental analysis of the inclusive $B^0$ photoproduction in $pPb$ and $PbPb$ collisions at the LHC is, in principle, feasible. In addition, that future experimental data will provide important constrains for the description of the hadronic structure at high energies, complementary to that derived from the study of exclusive processes.

\begin{figure}[t]
    \centering
    \includegraphics[width=0.49\linewidth]{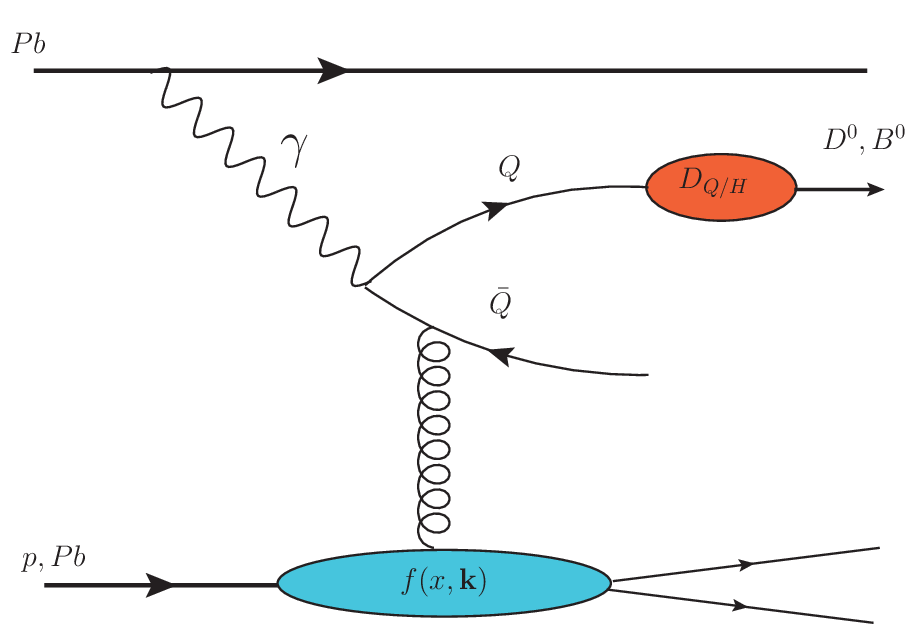}
    \caption{Inclusive heavy meson photoproduction in $pPb$ and $PbPb$ collisions.}
    \label{fig:diagram}
\end{figure}

This paper is organized as follows. In the next Section, we present a brief review of the formalism needed to describe the  inclusive heavy meson photoproduction in UPHICs. In particular, we will discuss the main ingredients used in our calculations. In Section \ref{sec:results} we will present our predictions for the rapidity and transverse momentum distributions associated with the inclusive $D^0$ and $B^0$ photoproduction in $pPb$ and $PbPb$ collisions at the LHC, derived considering  distinct models for the target unintegrated gluon distributions. Moreover, a comparison with the   CMS and preliminary ALICE data for the $D^0$ photoproduction will be performed. Predictions for the total cross-sections, derived considering the rapidity ranges covered by typical central and forward detectors, will also be presented. In addition, the contribution associated with the $b \rightarrow D^0$ transition will be estimated. Finally, in Section \ref{sec:summary} we will summarize our main conclusions.

\section{Formalism}
In this section we present a brief review of the color dipole $S$-matrix formalism, which was proposed in Refs. \cite{Nikolaev:2003zf,Nikolaev:2004cu,Nikolaev:2005dd,Nikolaev:2005zj,Nikolaev:2005ay,Nikolaev:2005qs}, and extended for the inclusive $D$ - meson photoproduction in $PbPb$ collisions in Ref.~\cite{Goncalves:2025wwt}. As  in Ref.~\cite{Goncalves:2025wwt}, we will assume that one of the hadrons as the photon source (hadron $A$) and the other as the target (hadron $B$) and that the hadron that emits the photon is moving in direction of positive
rapidities. Following Ref.~\cite{Goncalves:2025wwt},  the  differential cross - section for the inclusive photoproduction of a heavy - meson $H$ with transverse momentum $p_{T,H}$ at rapidity $Y_H$ in an ultraperipheral hadronic collision can be expressed as follows
\begin{eqnarray}
    \frac{d^2\sigma(A+B\rightarrow A\otimes H+X)}{dY_Hd^2p_{T,H}}=\int_{z_{min}}^1\frac{dz}{z^2}D_{Q/H}(z,\mu^2)\left[\frac{d^2\sigma(A+B\rightarrow A\otimes Q\Bar{Q}+X)}{dY_Qd^2p_T}\right]_{p_T=\frac{p_{T,H}}{z}}
    \label{eq:spectrumAB_meson}
\end{eqnarray}
where  $\otimes$ represents the presence of a rapidity gap in the final state, associated with the photon exchange, and 
$D_{Q/H}(z,\mu^2)$ is the fragmentation function for the heavy quark $Q$. 
Moreover, assuming the validity of the equivalent photon approximation,  the  cross - section for the production of a heavy - quark pair $Q\bar{Q}$ in a hadronic collision can be factorized in terms of the effective nuclear photon flux, $f^{\rm eff}_{\gamma/A}$, and the cross - section that describes the photon - hadron interaction, being given by  
\begin{eqnarray}
    \frac{d\sigma(A+B\rightarrow A\otimes Q\Bar{Q}+X)}{dY_Qd^2p_{T}}=\int_{x_{Q}}^1 {dz_Q} \frac{x_Q}{z_Q} f_{\gamma/A}^{\mathrm{eff}}\left(\frac{x_Q}{z_Q}\right)\frac{d\sigma(\gamma B\rightarrow Q\Bar{Q}X)}{dz_Qd^2p_T} \,\,,
\label{eq:spectrumAB_HQ}
\end{eqnarray}
where $x_Q=\sqrt{{(p_T^2+m_Q^2)}/{s_{NN}}}\exp(+Y_Q)$, $m_Q$ is the quark mass and $\sqrt{s_{NN}}$ is the center - of - mass energy of the $AB$ collision. In the color dipole $S$-matrix formalism, we have that~\cite{Goncalves:2025wwt}
\begin{align}
    \frac{d\sigma_T(\gamma B\rightarrow Q\bar{Q}X)}{dz_Qd^2{p_T}}=\frac{N_c\alpha_{em}e_Q^2}{2\pi^2}&\int d^2\mathbf{k}f_B(x,\mathbf{k})\left\{\left[z_Q^2+(1-z_Q)^2\right]\mathcal{B}_1(\mathbf{p}_T,\mathbf{k})+m_Q^2\mathcal{B}_2(\mathbf{p}_T,\mathbf{k})\right\} \,\,,
\end{align}
with $f_B(x,\mathbf{k})=({4\pi\alpha_s}/{3k^2})\mathcal{F}_B(x,\mathbf{k})$, where  $\mathcal{F}_B(x,\mathbf{k})$ is the unintegrated gluon distribution (UGD) of hadron $B$, which is determined by the QCD dynamics. Moreover, the auxiliary functions $\mathcal{B}_1$ and $\mathcal{B}_2$ are defined by
\begin{align}
    \mathcal{B}_1(\mathbf{p}_T,\mathbf{k})=&\frac{1}{2}\left[\frac{\mathbf{p}_T}{p^2_T+m_Q^2}-\frac{\mathbf{p}_T+\mathbf{k}}{(\mathbf{p}_T+\mathbf{k})^2+m_Q^2}\right]^2 \,\,,\\
    \mathcal{B}_2(\mathbf{p}_T,\mathbf{k})=&\frac{1}{2}\left[\frac{1}{p^2_T+m_Q^2}-\frac{1}{(\mathbf{p}_T+\mathbf{k})^2+m_Q^2}\right]^2\,\,.
\end{align}

In our analysis, we will consider ultraperipheral collisions, which are characterized by an impact parameter larger than the sum of the radius of incident hadrons ($b > R_A + R_B$). Moreover, we will to take into account the possibility of additinal soft electromagnetic interactions between the incident hadrons which can lead to breakup of the photon - emitting hadron. As a consequence, the effective nuclear photon flux will be given by 
 \begin{eqnarray}
x f^{\rm eff}_{\gamma/A}(x) = 
 \int d^2\rb \, \omega N_{A}(\omega,\rb)P(\rb) P_{\rm strong}(\rb)    \,\,,
 \end{eqnarray}
where $x = 2 \omega/\sqrt{s_{NN}}$, with $\omega$ being the photon energy and $P_{\rm strong}(\rb)$ and $P(\rb)$ representing the ``survival probabilities'' against strong interactions and electromagnetic dissociation,
respectively. In particular, $P_{\rm strong}(\rb)$ excludes configurations where hadrons come into contact, and is approximated by us as $  P_{\rm strong}(\rb) = \theta(b -  R_{A} - R_B), b = |\rb|$. On the other hand, $P(\rb)$ will be expressed by
\begin{eqnarray}
    P(\rb)  = \exp[ - \frac{S}{b^2}] \, , \quad S =  \frac{Z^2 \alpha_{\rm em}}{\pi^2}\int^\infty_{E_{\rm min}}  \frac{dE}{E} \sigma_{\rm tot} (\gamma A; E) \, ,
\end{eqnarray}
where $\sigma_{\rm tot} (\gamma A; E)$ is the total photoabsorption cross-section and $E$ denotes the photon energy in the target rest frame (See Refs.~\cite{Goncalves:2025wwt,Klusek-Gawenda:2013ema} for a detailed discussion).
Finally, we will assume the photon spectrum for the pointlike source, 
which reads \cite{upc}
\begin{eqnarray}
    \omega N_A(\omega,\rb) = \frac{Z^2 \alpha_{\rm em}}{\pi^2} \, \frac{1}{b^2} \, \, \, \left( \frac{\omega b}{\gamma} \right)^2 K_1^2 \Big(\frac{\omega b}{\gamma}\Big) \, , 
\end{eqnarray}
where $\alpha_{\rm em}$ is the electromagnetic fine structure constant and $\gamma$ is the Lorentz factor,  given by $\gamma = \gamma_{\rm cm} \equiv \sqrt{s_{NN}}/(2m_N)$. It is important to emphasize that for $P(\rb)=1$, one obtains the well-known result \cite{upc} 
\begin{equation}
    \label{fluxonucleo}
    f^{\rm eff}_{\gamma/A}(x)= \frac{2Z^2\alpha_{\rm em}}{\pi x}\left [\xi K_0(\xi)K_1(\xi)-\frac{\xi^2}{2}\left ( K_1^2(\xi)-K_0^2(\xi) \right )  \right ],
\end{equation}
where $\xi=(R_{A}+R_B)xm_N$ and the functions $K_i$ are the modified Bessel functions.

The main inputs to estimate the inclusive heavy meson photoproduction in hadronic collisions are the unintegrated gluon distribution of the hadron target, $\mathcal{F}(x,\mathbf{k})$, and the heavy quark fragmentation function, $D_{Q/H}$. The modeling of the fragmentation function will be discussed in the next section. For   $\mathcal{F}(x,\mathbf{k})$,  as in Ref.~\cite{Goncalves:2025wwt}, we will consider models based on the distinct assumptions for the QCD dynamics. In particular, for a proton target, we will consider the CCFM parametrization (setA1), derived in Ref. \cite{Jung:2004gs}  solving numerically the Ciafaloni - Catani - Fiorani - Marchesini (CCFM equation) \cite{Ciafaloni:1987ur,Catani:1989yc,Catani:1989sg} via Monte Carlo method, with the free parameters adjusted in order to describe the HERA data. It is important to emphasize that such an equation 
 is almost equivalent to BFKL equation in the regime of asymptotic energies and similar to the
DGLAP evolution for large values of $x$ and high values of the hard scale. In addition, we will consider two distinct solutions of the running coupling Balitsky - Kovchegov (rcBK) equation \cite{Balitsky:1995ub,Kovchegov:1999yj}, which takes into account of nonlinear effects in the   QCD dynamics, obtained in Refs. \cite{ALbacete:2010ad,Albacete:2012xq} and \cite{Kutak:2012rf}, which we will denote rcBK and KS nonlinear, respectively. Finally, we also will estimate the cross-sections using the linear solution of the rcBK equation obtained in Ref.~\cite{Kutak:2012rf}, which disregards the nonlinear effects, which we will denote KS linear. On the other hand, for a nuclear target, we will consider the CCFM parameterization rescaled for a Lead ion, disregarding nuclear effects, as well 
the PB-EPPS16 parameterization,  which was proposed in Ref. \cite{Blanco:2019qbm} considering the parton branching approach and taking into account of the nuclear effects, as predicted by the EPPS16 parametrization \cite{Eskola:2016oht}. Both parameterizations are derived assuming that the QCD dynamics is described by a linear evolution equation. Finally, we also will consider the nuclear UGD derived in Refs. \cite{ALbacete:2010ad,Albacete:2012xq} by solving the rcBK evolution equation. The comparison between the predictions of these distinct models will allow us to investigate the impact of the nuclear and nonlinear effects on the heavy meson photoproduction at the LHC energies.

\section{Results}
\label{sec:results}
In this section, we will present our predictions for the inclusive $D^0$ and $B^0$ photoproduction in $pPb$ and $PbPb$ collisions using the color dipole $S$ - matrix framework and distinct models for the proton and nuclear UGD's. In our analysis, we will assume $m_c=1.4$ GeV and $m_b=4.5$ GeV, and consider $pPb$ ($PbPb$) collisions at $\sqrt{s} = 8.1$ TeV ($\sqrt{s} = 5.36$ TeV). In what follows, we will initially discuss in the subsection \ref{subsec:frag} the impact of the modeling of the fragmentation functions (FF's) on our predictions. In particular, we will investigate the modifications on the predictions derived in Ref.~\cite{Goncalves:2025wwt} using the Peterson fragmentation function \cite{Peterson:1982ak}, when we instead consider a FF that evolves with the hard scale and is a solution of the DGLAP evolution equation.  
In the subsection \ref{subsec:PbPb} we will present our results for PbPb collisions, presenting a comparison with the CMS and preliminary ALICE data for the $D^0$ photoproduction. Moreover, predictions for the $B^0$ photoproduction are presented for the first time. In subsection \ref{subsec:pPb}, we investigate the inclusive heavy meson photoproduction in $pPb$ collisions and present predictions for the transverse momentum and rapiditiy distributions. Moreover, we present our results for the total cross-sections, derived  assuming the typical rapidity ranges covered by central and forward detectors at the LHC. Finally, in the subsection \ref{subsec:feed} we estimate the contribution for the $D^0$ photoproduction associated with the $b \rightarrow D^0$ transition.

\begin{figure}[t]
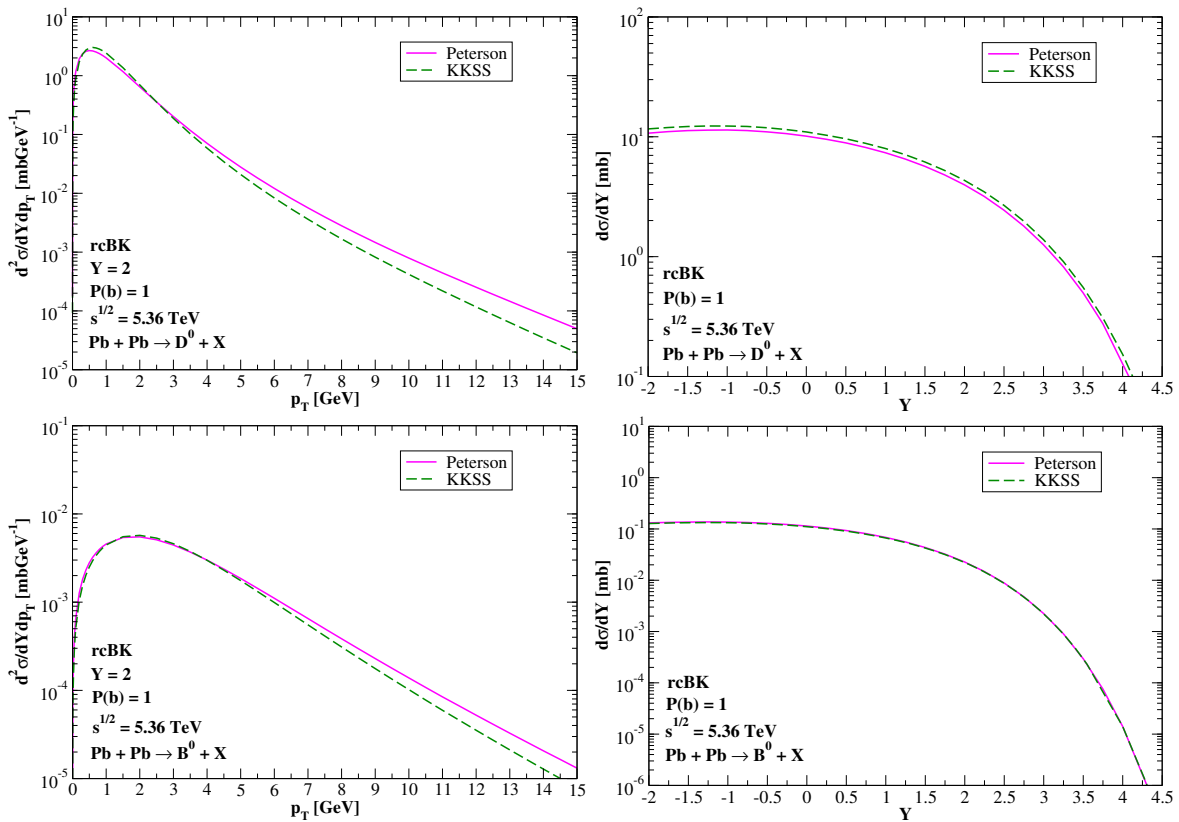

    \centering
    \includegraphics[width=0.43\linewidth]{spectra_p_Y_rcBK_Y_2_comparacao.eps}
    \includegraphics[width=0.43\linewidth]{rapidity_PbPb_comparacao.eps}
    \includegraphics[width=0.43\linewidth]{spectra_p_PbPb_Y_2_comparacao.eps}
    \includegraphics[width=0.43\linewidth]{rapidity_B0_PbPb_5.36_comparacao.eps}
    \caption{Impact of the fragmentation function modeling on the transverse momentum (left panel) and rapidity (right panels) distributions for the inclusive $D^0$ (upper panels) and $B^0$ (lower panels) photoproduction in $PbPb$ collisions at $\sqrt{s} = 5.36$ TeV. } 
        \label{fig:fragdependence}
\end{figure}

\subsection{Dependence on the fragmentation function}
\label{subsec:frag}
As discussed in the previous section, the predictions for the inclusive heavy meson photoproduction are dependent on the modeling of the fragmentation functions, $D_{Q/H} (z,\mu^2)$, which describes the hadronization of the heavy quark $Q$ into a meson $H$ with an energy fraction $z = E_H/E_Q$ at the hard scale $\mu^2$. 
 One of the simplest models present in the literature was proposed in Ref.~ \cite{Peterson:1982ak}, and is usually denoted Peterson fragmentation function. In this model, the fragmentation function is independent of hard scale $\mu^2$, and is given by
\begin{equation}
    D^{Q/H}(z)=\frac{n(H)}{z\left(1-\frac{1}{z}-\frac{\epsilon_Q}{1-z}\right)^2}
    \label{peterson}
\end{equation}
where $n(H)$ is the normalization factor and $\epsilon_Q$ is a non-perturbative parameter that depends on the quark flavor and is roughly proportional to the inverse square of the quark mass. In our analysis, we will assume $\epsilon_c=0.05$, $\epsilon_b=0.006$, $n(D^0)=0.308377$, and $n(B^0)=0.0528288$.
Another possibility is to obtain the  fragmentation functions by performing a global analysis of the existing data for  
heavy meson production, e.g. in $e^+ e^-$  or $pp$ collisions, using the DGLAP evolution equations for a given perturbative order. As a consequence, distinctly from the Peterson model, in this approach the FF depends on the hard scale. We have that once the FF is determined at some initial scale $\mu_0$, their evolution for another scale $\mu$ is determined by the DGLAP evolution equations. Over the last decades, several groups have performed this global analysis and derived distinct parameterizations for the heavy meson fragmentation functions (See, e.g., Refs.~\cite{D0_fragmentation,B0_fragmentation,Kniehl:2011bk,Kniehl:2012ti,MoosaviNejad:2013ssd,Kniehl:2015fla,Soleymaninia:2017xhc,Salajegheh:2019ach,Salajegheh:2019nea}). In our analysis, we will consider the parameterizations proposed in Refs. \cite{Kniehl:2012ti,Kniehl:2015fla}, which we will denote KKSS hereafter.  { For the hard scale, we choose $\mu^2 = p_{T,H}^2 + m_H^2$.}

In Fig. \ref{fig:fragdependence} we estimate the transverse momentum (left panels) and rapidity (right panels) distributions associated with the inclusive $D^0$ (upper panels) and $B^0$  (lower panels) photoproduction in $PbPb$ collisions at $\sqrt{s} = 5.36$ TeV, considering distinct models for the fragmentation function. These results were derived assuming the rcBK nuclear UGD and that $P(\mathbf{b})=1$, but we have verified that similar results are obtained for others UGD's and magnitudes of $P(\mathbf{b})$. The left panels of Fig. \ref{fig:fragdependence} clearly demonstrates the importance of the onset of QCD evolution in the fragmentation functions, with the importance of the DGLAP evolution increasing with the transverse momentum $p_T$. We have that the KKSS model  gives rise
to a suppression of heavy mesons at large $p_T$ compared to the Peterson result. On the other hand, the right panels indicate that the rapidity distributions are almost insensitive to the modeling of the fragmentation functions, which is expected, since these distributions are dominated by the contribution of small values of $p_T$, where the predictions from these two FF models are similar.

\begin{figure}[t]
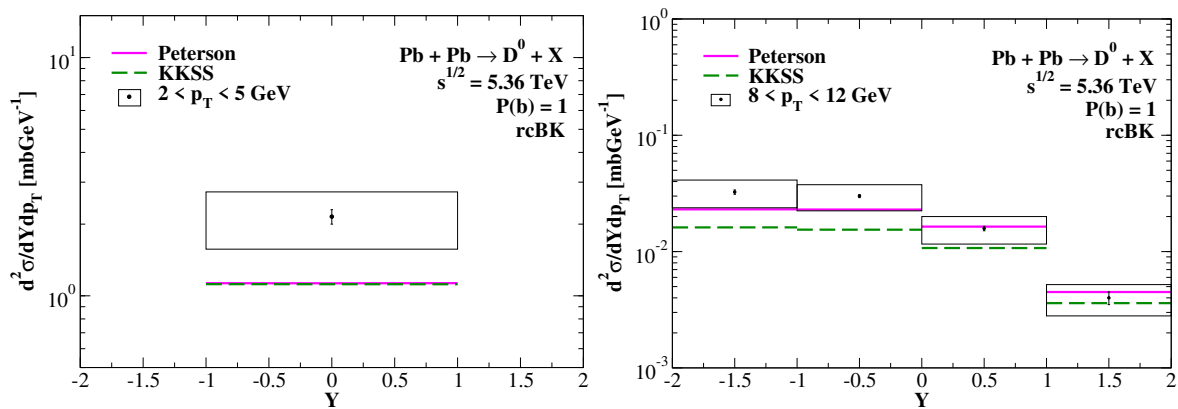

       \includegraphics[width=0.43\linewidth]{dsigmadYdp_rcBK_1_comparacao.eps}
    \includegraphics[width=0.43\linewidth]{dsigmadYdp_rcBK_3_comparacao.eps}
    \caption{Comparison between the predictions derived assuming the  Peterson and KKSS FF models  and the recent CMS data \cite{CMS:2025jjx}.}
    \label{fig:frag_cms}
\end{figure}

In Fig.~\ref{fig:frag_cms} we present a comparison of our predictions with the recent  CMS data \cite{CMS:2025jjx} for the rapidity distribution considering different transverse momentum bins associated with the inclusive $D^0$ photoproduction in ultraperipheral $PbPb$ collisions at $\sqrt{s} = 5.36$ TeV.
These data are for  the average differential distribution, defined by
\begin{eqnarray}
	\left\langle \frac{d^2 \sigma(Pb + Pb \rightarrow Pb \otimes D^0 + X)}{dY_D d^2p_{T,D}} \right\rangle = \frac{1}{\Delta p_T} \frac{1}{\Delta y} \int_{y_{min}}^{y_{max}} dy  \int_{p_T^{min}}^{p_{T}^{max}}  d p_{T,D} \frac{d^2 \sigma(Pb + Pb \rightarrow Pb \otimes D^0 + X)}{dY_D d^2p_{T,D}}  \,\,,
\end{eqnarray}
with $\Delta p_T = p_{T}^{max} - p_T^{min}$ and $\Delta y = y_{max} - y_{min}$, where $p_T^{max}$ ($y_{max}$)  and $p_T^{min}$ ($y_{min}$) are the upper and lower values of the transverse momentum (rapidity) in a given bin. 
We have that for the bin associated with smaller values of $p_T$, the predictions are almost identical, but differ when larger values of $p_T$ are considered, as expected from our previous analysis. In particular, the inclusion of evolution in the FF implies that  the that the description of the data for negative rapidities, associated with the calculations performed for this combination of UGD + $P(\mathbf{b})$, is deteriorated.

\begin{figure}[t]
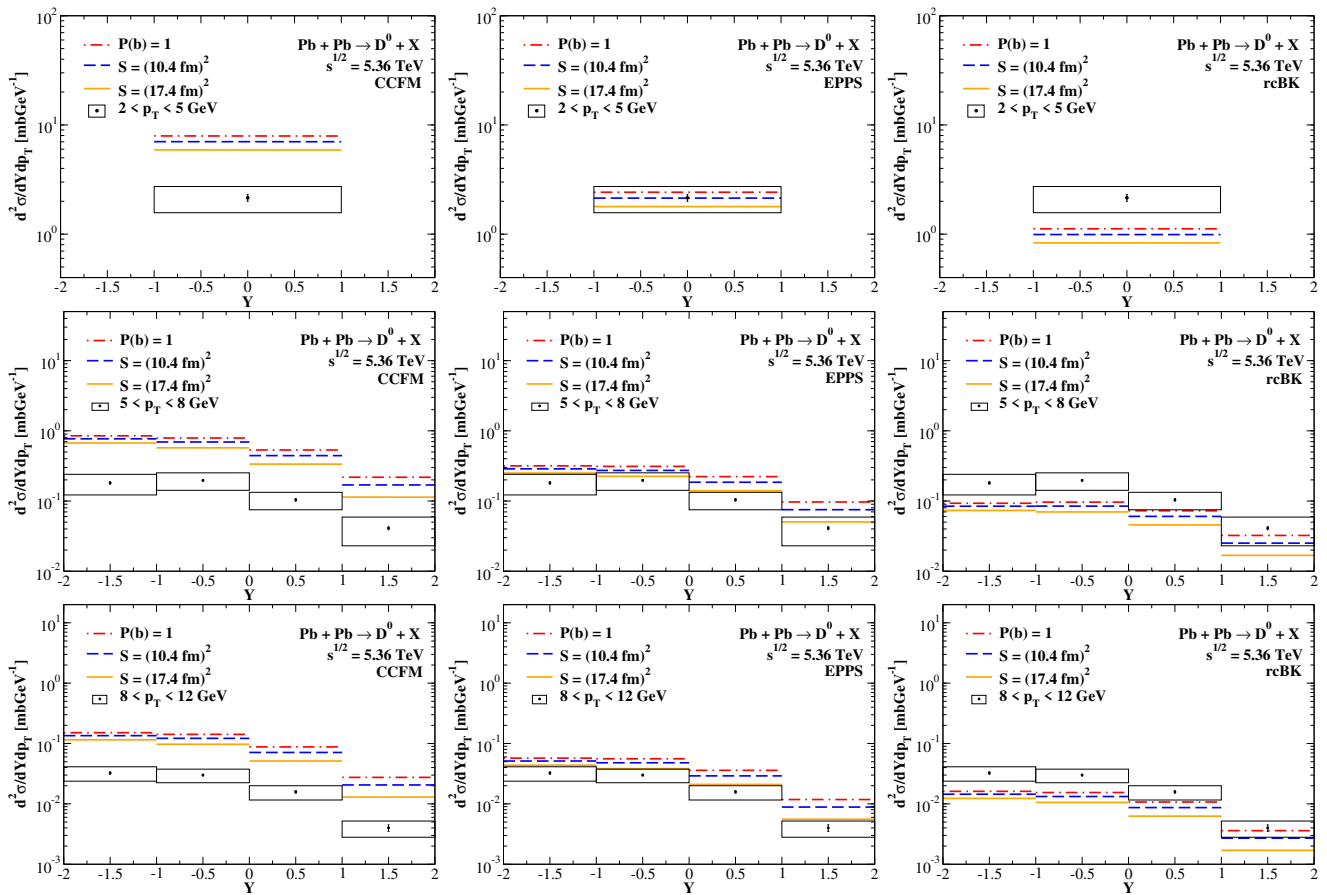

    \centering
    \includegraphics[width=0.32\linewidth]{dsigmadYdp_CCFM_1.eps}
    \includegraphics[width=0.32\linewidth]{dsigmadYdp_EPPS_1.eps}
    \includegraphics[width=0.32\linewidth]{dsigmadYdp_rcBK_1.eps}
    \includegraphics[width=0.32\linewidth]{dsigmadYdp_CCFM_2.eps}
    \includegraphics[width=0.32\linewidth]{dsigmadYdp_EPPS_2.eps}
    \includegraphics[width=0.32\linewidth]{dsigmadYdp_rcBK_2.eps}
    \includegraphics[width=0.32\linewidth]{dsigmadYdp_CCFM_3.eps}
    \includegraphics[width=0.32\linewidth]{dsigmadYdp_EPPS_3.eps}
    \includegraphics[width=0.32\linewidth]{dsigmadYdp_rcBK_3.eps}
    \caption{Rapidity distribution associated with the inclusive $D^0$ photoproduction in ultraperipheral $PbPb$ collisions at $\sqrt{s}=5.36$ TeV. Data from CMS collaboration \cite{CMS:2025jjx} obtained considering distint bins in the meson transverse momentum. Theoretical predictions were derived considering distinct models for the nuclear UGD and for the description of the electromagnetic dissociation. }
    \label{fig:cms_data}
\end{figure}

\subsection{$PbPb$ collisions}
\label{subsec:PbPb}
In this subsection we will present our predictions for the inclusive heavy meson photoproduction in $PbPb$ collisions. Hereafter, we will estimate the cross-sections assuming the KKSS model for the fragmentation function, which takes into account of the DGLAP evolution. In particular, in Fig.~\ref{fig:cms_data} we update the predictions for the average rapidity distributions presented in Ref.~\cite{Goncalves:2025wwt}, by considering this new FF model instead of the Peterson fragmentation function. The predictions have been considering distinct models for the nuclear UGD and for the description of the electromagnetic dissociation. 
One has that the CCFM prediction, which disregard nuclear effects, is not able to describe the data, as already verified in  Ref.~\cite{Goncalves:2025wwt}. In contrast, the description of data by the EPPS (rcBK) model, which includes nuclear effects, is improved (deteriorated). Unfortunately, more definite conclusions are not yet possible. As already pointed out in Ref.~\cite{Goncalves:2025wwt}, more precise data, derived considering smaller bins in $p_T$ and a larger range of rapidity would be very useful to improve our description of the nucleus at small - $x$.

\begin{figure}[t]
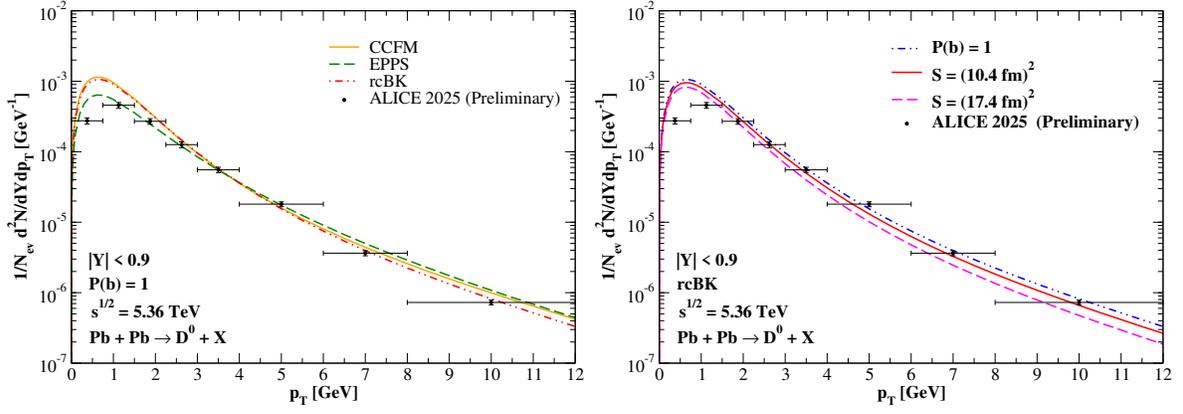

    \centering
    \includegraphics[width=0.43\linewidth]{pT_D0_PbPb_5.36_ALICE_KKKS.eps}
    \includegraphics[width=0.43\linewidth]{pT_S_D0_PbPb_5.36_ALICE_KKKS.eps}
    \caption{Transverse momentum distributions for the inclusive $D^0$ photoproduction in $PbPb$ collisions. Preliminary data from ALICE Collaboration \cite{Nese:2025ohz}. Dependencies on the nuclear UGD (left panel) and on the description of the electromagnetic dissociation (right panel). }
    \label{fig:alice}
\end{figure}

The ALICE Collaboration is currently also performing a detailed analysis of the inclusive $D^0$ photoproduction in $PbPb$ collisions \cite{ALICE} and has recently presented the first preliminary data for the transverse momentum distribution { for $|Y_D| < 0.9$} in Ref.~\cite{Nese:2025ohz}. In Fig.\ref{fig:alice} we compare our predictions, derived assuming different nuclear UGD's (left panel) or distinct assumptions for the treatment of the electromagnetic dissociation (right panel), with this preliminary data. Our predictions have been normalized as in Ref.~\cite{Nese:2025ohz}.  We have that the  current data for $p_T \gtrsim 2$ GeV is quite well described by the predictions associated with the distinct UGD's. In contrast, the data for smaller values of  $p_T$ is sensitive to the UGD, and is better described by the EPPS model. In addition, we have that the current data is not able to discriminate between the predictions based on distinct amounts for the electromagnetic dissociation (see right panel).

\begin{figure}[t]
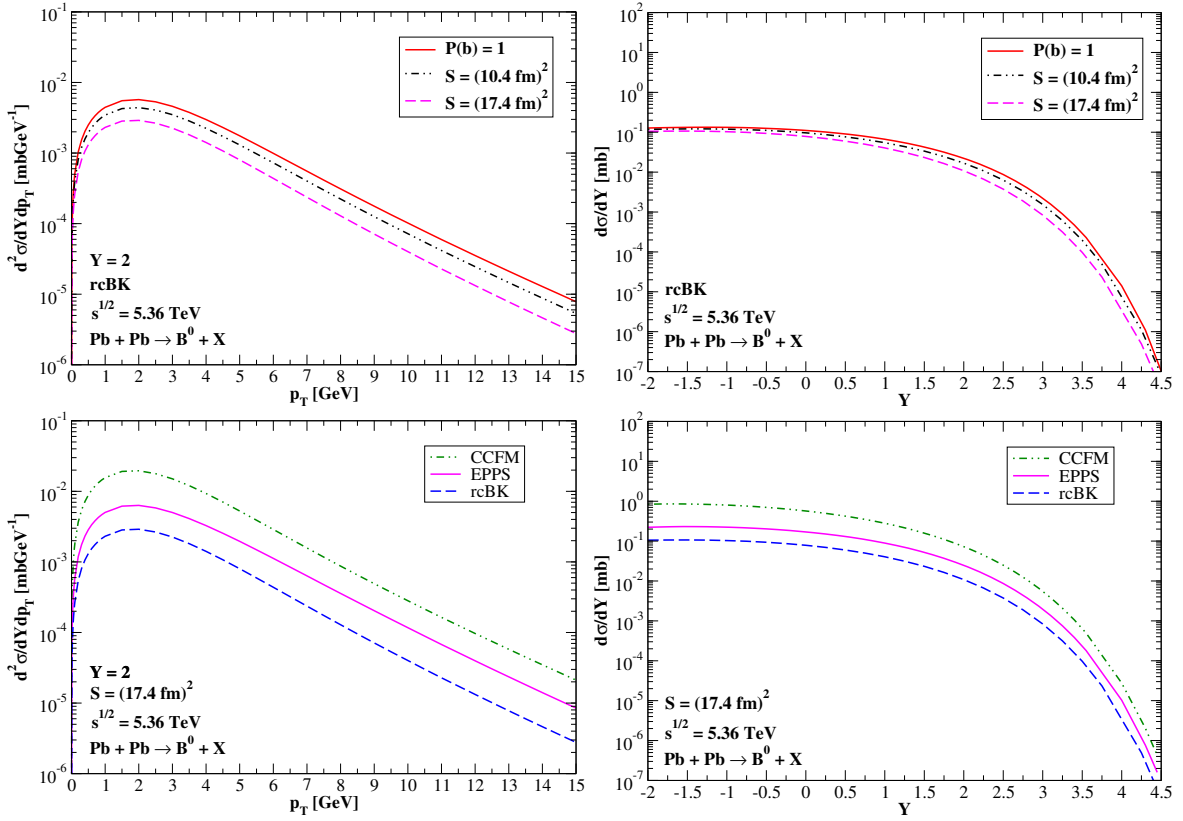

    \centering
    \includegraphics[width=0.43\linewidth]{spectra_p_PbPb_Y_2_rcBK.eps}
    \includegraphics[width=0.43\linewidth]{rapidity_B0_PbPb_5.36.eps}
        \includegraphics[width=0.43\linewidth]{pT_PbPb_Y_2_B0.eps}
    \includegraphics[width=0.43\linewidth]{rapidity_PbPb_B0.eps}

    \caption{Transverse momentum (left panels) and rapidity (right panels) distributions for the inclusive $B^0$ photoproduction in $PbPb$ collisions at $\sqrt{s} = 5.36$ TeV.  Dependencies on the  description of the electromagnetic dissociation (upper panels) and nuclear UGD's (lower panels). }
    \label{fig:B0_PbPb}
\end{figure}

In Fig.~\ref{fig:B0_PbPb} we present, for the first time, the predictions for the inclusive $B^0$ photoproduction in $PbPb$ collisions. The dependence on the treatment of the electromagnetic dissociation is presented in the upper panels. The calculations  have been performed  assuming the rcBK nuclear UGD, but we have verified that similar results are obtained  for the other two parametrizations, differing only in the normalization. The results indicate that, similarly to observed for $D^0$ photoproduction in Ref.~\cite{Goncalves:2025wwt}, the inclusion of the nuclear dissociation implies a decreasing of the normalization of the distributions, with the impact being larger with the increasing of the parameter $S$. In addition, the impact is also larger when the rapidity is increased. In the lower panels, we estimate the dependence on the nuclear UGD assumed in the calculation. We have that  the CCFM predictions have the larger normalization, which is expected, since this parameterization disregards the impact of nuclear and nonlinear effects. The inclusion of the nuclear effects, present in the EPPS parametrization, implies that the normalization is decreased in comparison to the CCFM one. Finally, the rcBK prediction has a smaller normalization. Such results indicate that a future experimental analysis of this final state can be useful to constrain the description of the QCD dynamics and/or the magnitude of the nuclear effects.

\subsection{$pPb$ collisions}
\label{subsec:pPb}

In this subsection, we will focus on the possibility of using the inclusive heavy meson photoproduction in $pPb$ collisions as a way to constrain the description of the QCD dynamics in a proton target. In what follows, the transverse momentum and rapidity distributions will be estimated assuming distinct proton UGD's, based on different assumption for the underlying dynamics. In particular, we will consider two UGD's that are solution of the rcBK evolution equation, denoted rcBK and KS nonlinear~\cite{ALbacete:2010ad,Albacete:2012xq,Jung:2004gs,Kutak:2012rf}, and will compare its predictions with those obtained disregarding the nonlinear effects (CCFM and KS linear)~\cite{Jung:2004gs,Kutak:2012rf}. In our calculations, we will assume $P(\mathbf{b})=1$, since the contribution of the electromagnetic dissociation in $pPb$ collisions is expected to be negligible. Moreover, we will consider that the nucleus that has emitted the photon is moving in direction of positive rapidities and the calculations will be performed in the center - of - mass frame.

\begin{figure}[t]
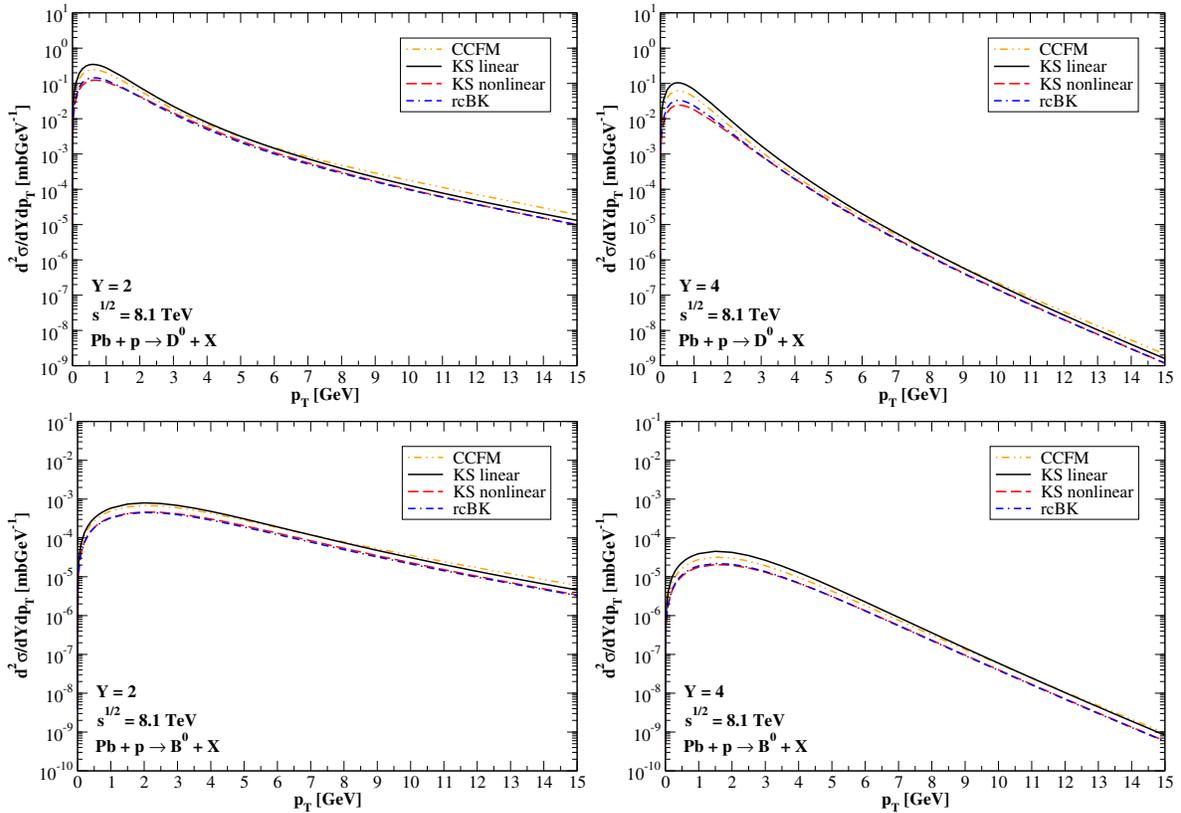

    \centering
    \includegraphics[width=0.43\linewidth]{pT_D0_Y_2_Pbp_8.1.eps}
    \includegraphics[width=0.43\linewidth]{pT_D0_Y_4_Pbp_8.1.eps}
    \includegraphics[width=0.43\linewidth]{pT_B0_Y_2_Pbp_8.1.eps}
    \includegraphics[width=0.43\linewidth]{pT_B0_Y_4_Pbp_8.1.eps}
    \caption{Transverse momentum distributions for the inclusive $D^0$ (upper panels) and $B^0$ (lower panels) photoproduction in $pPb$ collisions at $\sqrt{s} = 8.1$ TeV assuming two  distinct values for the meson rapidity: $Y = 2$ (left panels) and $Y = 4$ (right panels). Theoretical predictions for distinct proton UGD's.}
    \label{fig:pPb_trans}
\end{figure}

\begin{figure}[t]
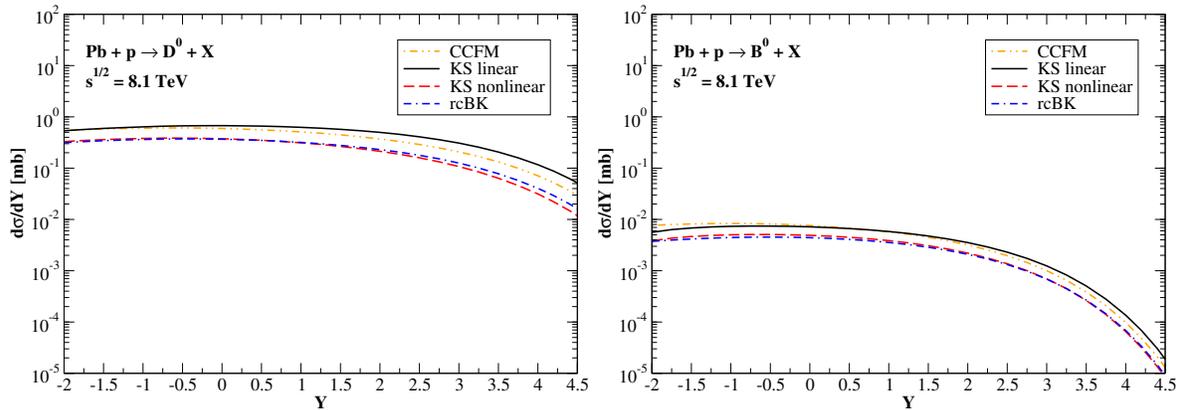

    \centering
    \includegraphics[width=0.43\linewidth]{rapidity_D0_Pbp_8.1.eps}
    \includegraphics[width=0.43\linewidth]{rapidity_B0_Pbp_8.1.eps}
    \caption{Rapidity distributions for the inclusive $D^0$ (left panels) and $B^0$ (right panels) photoproduction in $pPb$ collisions at $\sqrt{s} = 8.1$ TeV. Theoretical predictions for distinct proton UGD's.}
    \label{fig:pPb_rap}
\end{figure}

In Figs.~\ref{fig:pPb_trans} and ~\ref{fig:pPb_rap}, we present our predictions for the transverse momentum and rapidity distributions, respectively, associated with the inclusive  $D^0$  and $B^0$  photoproduction in $pPb$ collisions at $\sqrt{s} = 8.1$ TeV. These results indicate that the rcBK and KS  nonlinear predictions are quite similar, which is expected, since both are based on the rcBK evolution equation, differing only in the numerical implementation and assumptions for the initial condition. Such predictions provide the results with the smaller normalization. In contrast, when the nonlinear effects are disregarded, the associated predictions (CCFM and KS linear) have a larger normalization. Such results indicate that a future experimental analysis of these two final states can provide additional constrain on the description of the proton structure.

In Tables ~\ref{tab:PbPb} and ~\ref{tab:pPb} we present our predictions for the total cross-sections (in mb) for the heavy meson photoproduction in $PbPb$ collisions at $\sqrt{s}=5.36$ TeV and $pPb$ collisions at $\sqrt{s}=8.1$ TeV, respectively, derived considering the typical rapidity ranges covered by   central and forward detectors, which we assume to be $-2.0 \le Y \le +2.0$ and $2.0 \le Y \le 4.5$, respectively. Calculations for $PbPb$ collisions were performed assuming $S = (17.4 \, \mbox{fm})^2$. Our results indicate that the predictions for 
$PbPb$ collisions are approximately two orders of magnitude larger than those for $pPb$ collisions, which is partially associated with the $Z^2$-factor in the photon flux. We also have that the results for the $B^0$ photoproduction are smaller than for the $D^0$ case by a factor $\approx 10^2$, which is associated with the mass dependence of the cross-section. Moreover, the predictions decrease by one or two orders of magnitude when a forward rapidity selection is considered. Finally, we have that the results are  sensitive to the model assumed for the target UGD. Considering the typical luminosities for $PbPb$ and $pPb$ collisions at the LHC, ${\cal{O}}(nb^{-1})$, our results indicate that a future experimental analysis of the inclusive $B^0$ photoproduction is, in principle, feasible,  as well the measurement of the inclusive  $D^0$ photoproduction in $pPb$ collisions.

\begin{table}[h!]
    \centering
    \begin{tabular}{|c|c|c|}\hline
                        & $D^0$        & $B^0$    \\\hline
        CCFM            & 225.1 (14.69) & 2.138 (0.0310)\\
        EPPS            & 50.31 (3.717) & 0.615 (0.0107)\\
        rcBK            & 31.11 (1.867) & 0.284 (0.0046)\\
        \hline
            \end{tabular}
    \caption{Total cross-section (in mb) for the heavy meson photoproduction in $PbPb$ collisions at $\sqrt{s}=5.36$ TeV, derived considering the rapidity range covered by a typical central (forward) detector. Calculations performed for $S = (17.4 \, \mbox{fm})^2$.}
    \label{tab:PbPb}
\end{table}

\begin{table}[h!]
    \centering
    \begin{tabular}{|c|c|c|}\hline
                        & $D^0$        & $B^0$    \\\hline
        CCFM            & 2.18 (0.449) & 0.0261 (0.00241)\\
        KS linear       & 2.48 (0.656) & 0.0244 (0.00286)\\
        KS nonlinear    & 1.35 (0.236) & 0.0164 (0.00167)\\
        rcBK            & 1.33 (0.271) & 0.0149 (0.00162)\\
        \hline
    \end{tabular}
    \caption{Total cross-section (in mb) for the heavy meson photoproduction in $pPb$ collisions at $\sqrt{s}=8.1$ TeV, derived considering the rapidity range covered by a typical central (forward) detector.}
    \label{tab:pPb}
\end{table}

\subsection{Inclusive $D^0$ photoproduction from bottom hadron decays}
\label{subsec:feed}
Finally, in this subsection, we will investigate the contribution from bottom hadron decays for the inclusive $D^0$ photoproduction. In our analysis, we will consider the approach discussed in Ref.~\cite{Bolzoni:2013vya}, where this contribution is estimated considering the $b$-quark production and the posterior hadronization of this $b$ quark into a charmed meson. In this approach, we have a one-step transition based on the fragmentation function for $b \rightarrow D^0$. As a consequence, such a contribution can be easily implemented in our framework. We will calculate Eq.~(\ref{eq:spectrumAB_HQ}) for the $b$-quark production and use in the Eq.~(\ref{eq:spectrumAB_meson}) the FF for the $b \rightarrow D^0$ transition, provided in Ref.~\cite{Kniehl:2012ti}. 

\begin{figure}[t]
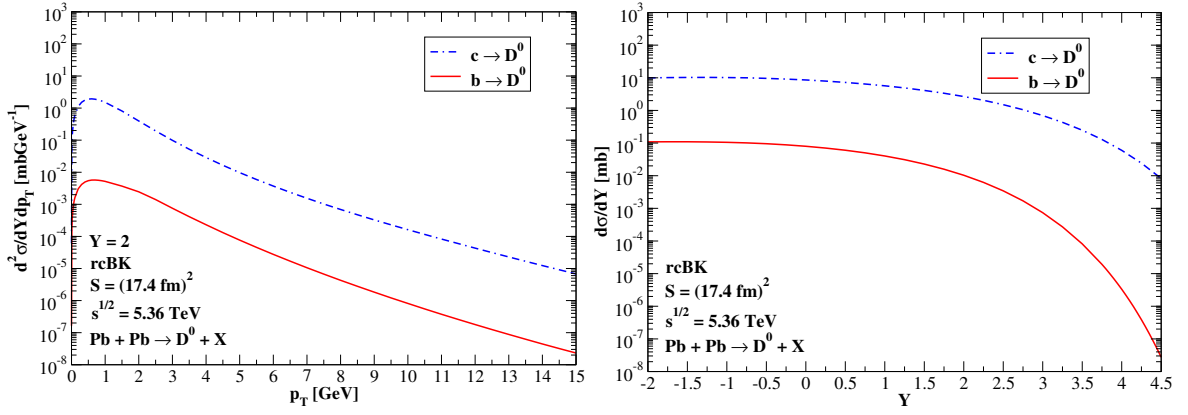

    \centering
        \includegraphics[width=0.43\linewidth]{pT_feed_down_PbPb_Y_2.eps}
     \includegraphics[width=0.43\linewidth]{rapidity_feed_down_PbPb_D0.eps}
    \caption{Transverse momentum (left panel) and rapidity (right panel) distributions for the inclusive $D^0$ photoproduction in $PbPb$ collisions at $\sqrt{s} = 5.36$ TeV. Predictions for the contribution from bottom hadron decays for the inclusive $D^0$ photoproduction (solid red lines). For completeness, the results for the case where the $D^0$ meson comes from the fragmentation of charm quarks, denoted $c \rightarrow D^0$ in the plot, is also presented.}
    \label{fig:fragbtoD}
\end{figure}

In Fig.~\ref{fig:fragbtoD} we present our predictions for the transverse momentum (left panel) and rapidity (right panel) distributions for the inclusive $D^0$ photoproduction in $PbPb$ collisions at $\sqrt{s} = 5.36$ TeV, associated with the contribution from bottom hadron decays for the inclusive $D^0$ photoproduction (solid red lines). For completeness, the results for the case where the $D^0$ meson comes from the fragmentation of charm quarks (dashed blue lines) is also presented. We have that the amount of $D^0$ mesons originating from $b$ quark decays is only a small fraction of the dominating contribution, which is associated with the fragmentation of charm quarks, in agreement with the results derived in Ref.~\cite{Bolzoni:2013vya}. Although  these results have been derived using the rcBK nuclear UGD and assuming $S = (17.4 \, \mbox{fm})^2$,  we have verified that the conclusions are also valid for other UGD's and treatments of $P(\mathbf{b})$, as well for $pPb$ collisions.

\section{Summary}
\label{sec:summary}
One of the main goals of Particle Physics is to achieve a deeper knowledge of the hadronic structure. Over the last decades, several studies have demonstrated that our understanding of the QCD dynamics at high energies can be significantly improved through the analysis of  photon-induced interactions present in electron-hadron and hadron-hadron collisions. In particular, the study of exclusive and inclusive processes in ultraperipheral $pPb$ and $PbPb$ collisions is expected to provide additional and complementary constrains on the description of proton and nuclear targets. In this paper, we have focused on the inclusive $D^0$ and $B^0$ photoproduction and estimated the associated transverse momentum  and rapidity distributions, as well total cross-sections, using the    color dipole $S$-matrix formalism  and considering different models for the target unintegrated gluon distribution, based on distinct assumptions for the description of the QCD dynamics. In particular, for a nuclear target, we have compared the predictions derived assuming (or not) the presence of nuclear effects with those based on the solution of the rcBK evolution equation, which takes into account of the nonlinear effects. A similar comparison between linear and nonlinear predictions have also been performed for a proton target. Such analysis has demonstrated that the inclusive heavy meson photoproduction can be used to constrain the description of the hadronic structure. In addition, our results have demonstrated that a future experimental of inclusive $D^0$ and $B^0$ photoproduction in $pPb$ collisions is, in principle, feasible, as well the $B^0$ photoproduction in $PbPb$ collisions at the LHC.

\begin{acknowledgments}
  V.P.G. and L. S.  were partially supported by CNPq, CAPES (Finance Code 001), FAPERGS and INCT-FNA (Process No. 408419/2024-5). The work of W.S. was partially supported by the Polish National Science Center Grant No. UMO-2023/49/B/ST2/03665.
\end{acknowledgments}

\bibliographystyle{unsrt}

\end{document}